\documentclass[aps,prl,twocolumn,superscriptaddress,letterpaper]{revtex4}

\pdfoutput=1
\usepackage{amsmath,amsfonts,amssymb,amsbsy,amscd,amsgen}
\usepackage{subfigure}
\usepackage{float}
\usepackage{float}
\usepackage{color,graphicx}

\newcommand{\reffig}[1]{Fig.\ \ref{#1}}
\newcommand{\refFig}[1]{Fig.\ \ref{#1}}
\newcommand{\bu}{\ensuremath{{\bf u}}}
\newcommand{\bx}{\ensuremath{{\bf x}}}
\newcommand{\butw}{\ensuremath{{\bf u}_{\text{TW}}}}
\newcommand{\bueq}{\ensuremath{{\bf u}_{\text{EQ}}}}
\newcommand{\buxavg}{\ensuremath{\langle {\bf u} \rangle}_x}
\newcommand{\Reynolds}{\ensuremath{\mathrm{Re}}}
\newcommand{\colorswitch}[2]{#1}

\begin{document}

\title{Snakes and ladders: localized solutions of plane Couette flow}

\author{Tobias M. Schneider}
\affiliation{School of Engineering and Applied Science, Harvard University}

\author{John F. Gibson}
\affiliation{School of Physics, Georgia Institute of Technology}

\author{John Burke}
\affiliation{Department of Mathematics, Boston University}

\date{\today}

\begin{abstract}
We demonstrate the existence of a large number of exact solutions 
of plane Couette flow, which share the topology of known periodic
solutions but are localized in space. Solutions of different size are organized 
in a \emph{snakes-and-ladders} structure strikingly similar to that observed 
for simpler pattern-forming PDE systems. 
These new solutions are a step towards extending the dynamical systems 
view of transitional turbulence to spatially extended flows.
\end{abstract}



\maketitle

The discovery of exact equilibrium and traveling-wave solutions 
to the full nonlinear Navier-Stokes equations has resulted in much recent 
progress in understanding the dynamics of linearly stable shear flows such 
as pipe, channel and plane Couette flow~\cite{Nagata1990,WaleffeJFM01,Faisst2003,Wedin2004}.
These \emph{exact solutions}, together with their entangled stable and 
unstable manifolds, form a dynamical network that supports
chaotic dynamics, so that turbulence can be understood as a
walk among unstable solutions~\cite{LandfordARFM82,Gibson2008}. Moreover, 
specific exact solutions are found to be \emph{edge states}~\cite{Schneider2008},
that is, solutions with codimension-1 stable manifolds that locally form the 
stability boundary between laminar and turbulent dynamics. Thus, exact 
solutions play a key role both in supporting turbulence and in guiding 
transition.

This emerging dynamical systems viewpoint does not yet capture the full 
spatio-temporal dynamics of turbulent flows. One major limitation is 
that exact solutions have mostly been studied in small computational 
domains with periodic boundary conditions. The small periodic solutions 
cannot capture the localized structures typically observed in spatially 
extended flows. For example, pipe flows exhibit localized turbulent puffs.
Similarly, in plane Couette flow (PCF), the flow between two parallel 
walls moving in opposite directions, localized perturbations trigger 
turbulent spots which then invade the surrounding laminar 
flow~\cite{Tillmark1992,DaviaudPRL92}. 
Even more regular long-wavelength spatial patterns such as turbulent 
stripes have been observed~\cite{Barkley2005}.
The known periodic exact solutions cannot capture this rich spatial
structure, but they do suggest that {\em localized} solutions 
might be key in understanding the dynamics of spatially extended flows.

Spatially localized states are common in a variety of driven
dissipative systems. These are often found in a parameter regime of
bistability (or at least coexistence) between a spatially uniform
state and a spatially periodic pattern, such as occurs in a
subcritical pattern forming instability. The localized state then
resembles a slug of the pattern embedded in the uniform background. An
early explanation of such states is due to Pomeau~\cite{Pomeau1986},
who argued that a front between a spatially uniform and spatially
periodic state, which might otherwise be expected to drift in time,
can be stabilized over a finite parameter range by pinning to the spatial
phase of the pattern. More recently, the details of this localization 
mechanism have been established for the subcritical Swift-Hohenberg
equation (SHE) through a theory of {\em spatial 
dynamics}~\cite{Champneys98,Burke:2007p421,Chapman:2009p455}. In
one spatial dimension the time-independent version of this PDE can be
treated as a dynamical system in space, in which stationary profiles are
seen as trajectories in the spatial coordinate. Then localized
states correspond to homoclinic orbits to a fixed point that visit
the neighborhood of a periodic orbit representing the pattern. The SHE
is equivariant under spatial reflections, so the corresponding spatial
dynamical system is reversible. There exists an infinite multiplicity
of reversible homoclinic orbits (i.e., symmetric localized states)
organized in a pair of solution branches which undergo \emph{homoclinic
snaking}. In a bifurcation diagram the two branches intertwine,
oscillating back and forth within a parameter regime called the
snaking or pinning region. These are connected by branches of
non-symmetric states called \emph{rungs}. Together they form the
\emph{snakes-and-ladders} structure of localized states.

The theory of spatial dynamics also applies to other equations in 
one spatial dimension~\cite{MargaretBeck:2009p2216}, but there is no obvious
extension to higher dimensional PDEs. Nevertheless, there are
remarkable similarities between localized states in the simple
one-dimensional SHE and in other more realistic (and complicated)
PDEs. 
In fluid dynamics, homoclinic snaking occurs in driven 
two-dimensional systems such as binary fluid 
convection~\cite{Batiste:2006p539} and natural doubly diffusive 
convection~\cite{Bergeon:2008p530}. Localized solutions in these 
systems exhibit snaking in bifurcation diagrams and are homoclinic 
in that they transition along one of the spatial coordinates from a 
uniform state, to a periodic pattern, and back to the uniform state.
In shear flows, homoclinic snaking
has never been observed but its existence has been
speculated~\cite{Knobloch:2008p529}. This speculation is supported by
the recent discovery of two localized exact solutions in
PCF by Schneider, Marinc and Eckhardt~\cite{Schneider2009} which
qualitatively resemble localized states in the SHE.

The aim of this letter is to elucidate the origin of these localized
solutions in PCF. We show that the Navier-Stokes equations in this 
geometry indeed exhibit homoclinic snaking, giving rise to localized
counterparts of well-known spatially periodic equilibria.

\begin{figure}
{\small (a)} \includegraphics[width=0.45\textwidth]{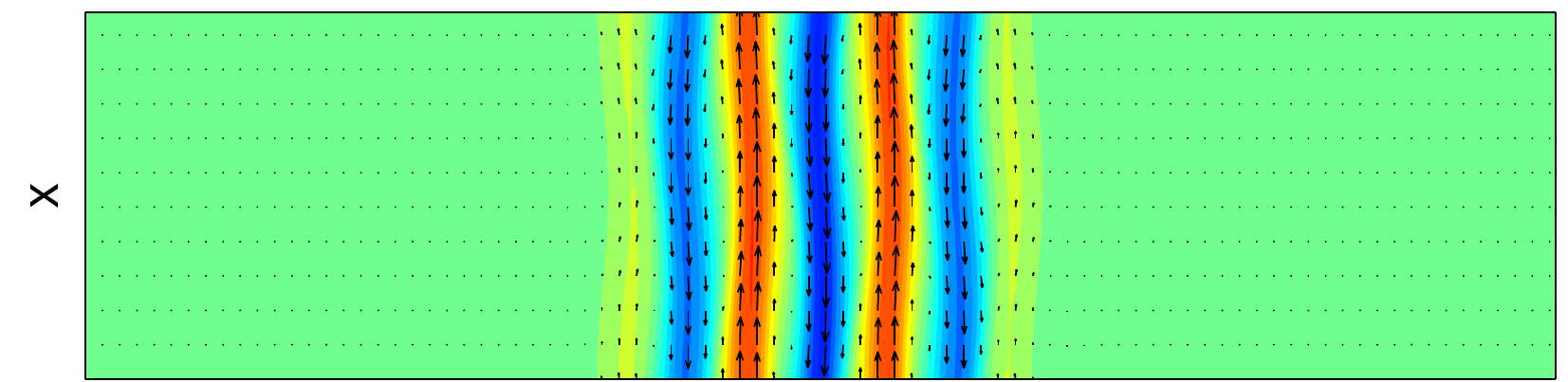} 
{\small (b)} \includegraphics[width=0.45\textwidth]{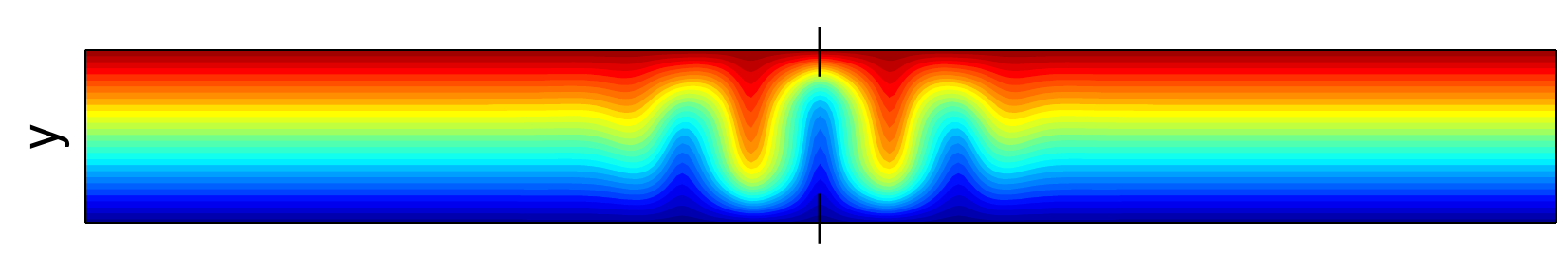} 
{\small (c)} \includegraphics[width=0.45\textwidth]{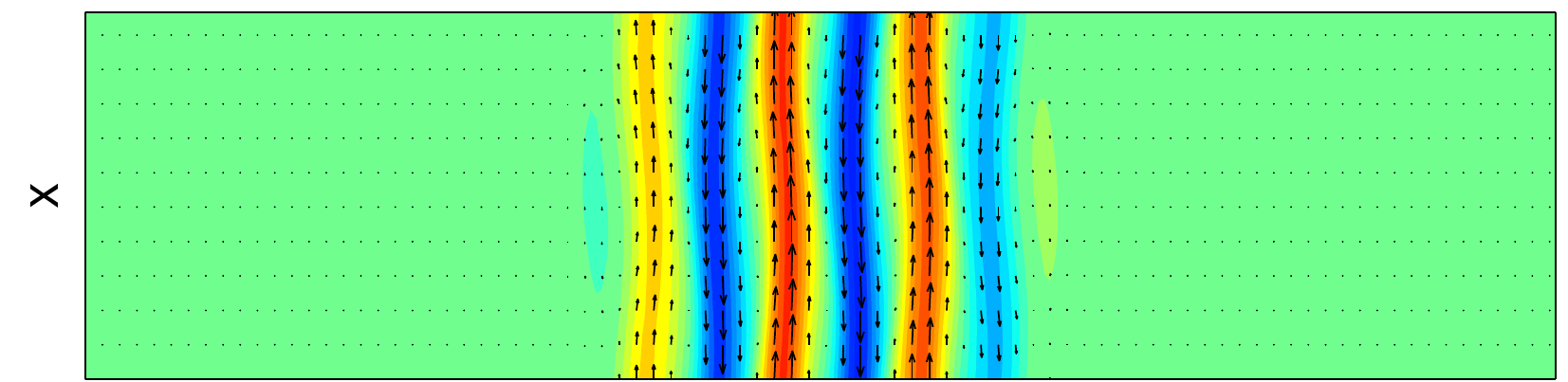} 
{\small (d)} \includegraphics[width=0.45\textwidth]{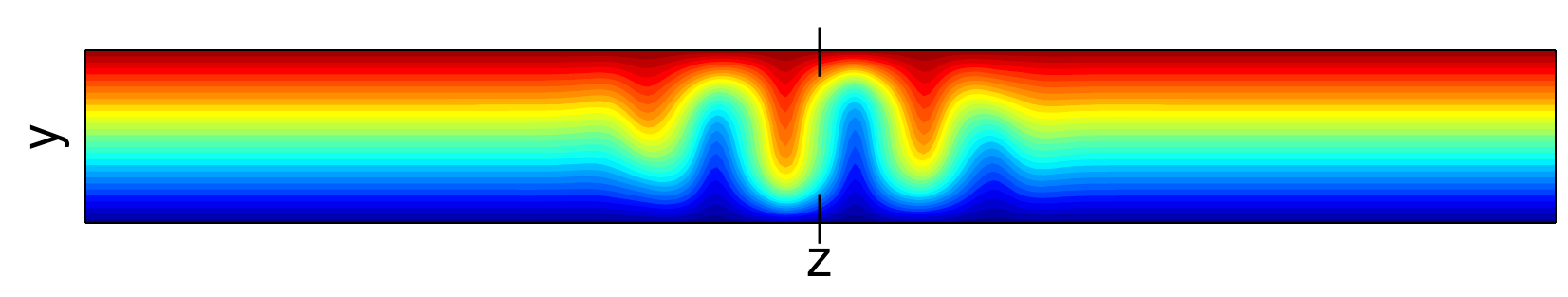} 
\caption{\label{fig:Re400fields}
(color online). Localized traveling wave $\butw$ (a,b) and equilibrium 
$\bueq$ (c,d) solutions of plane Couette flow at $\Reynolds=400$, from 
\cite{Schneider2009}. The velocity fields are shown in the $y=0$ midplane 
in (a,c), with arrows indicating in-plane velocity and the  
\colorswitch{color}{gray} scale indicating streamwise velocity $u$:  
\colorswitch{blue/green/red}{black/grey/white} correspond to $u=-1,0,+1$.
The $x$-averaged streamwise velocity is shown in (b,d), with $y$ expanded
by a factor of three.
}
\end{figure}

In PCF the velocity field $\bu(\bx,t) = [u,v,w](x,y,z,t)$
evolves under the incompressible Navier-Stokes equations,
\begin{equation} 
\frac{\partial \bu}{\partial t} + \bu \cdot \nabla \bu = -\nabla p + \frac {1} {\Reynolds} \nabla^2 \bu,\quad \nabla \cdot \bu = 0,
\label{eq:NSE}
\end{equation}
in the domain $\Omega = L_x\!\times\!L_y\!\times\!L_z$ where $x$, $y$, $z$ are the 
streamwise, wall-normal, and spanwise directions, respectively. 
The boundary conditions are periodic in $x$ and $z$ and no-slip at the walls, 
$\bu(y=\pm 1) = \pm {\bf \hat{x}}$. The Reynolds number is $\Reynolds=Uh/\nu$, 
where $U$ is half the relative velocity of the walls, $h$ half the wall separation,
and $\nu$ the kinematic viscosity. We treat $\Reynolds$ as the control parameter
and use as a solution measure the dissipation rate 
$D = (L_x L_y L_z)^{-1} \int_{\Omega} (|{\bf \nabla} \times \bu|^2 ) \,\mathrm{d}\Omega$.
The laminar profile has $D=1$ while solutions such as those shown in 
Fig.~\ref{fig:Re400fields} have $ D>1$.

Figure~\ref{fig:Re400fields} shows two exact solutions of (\ref{eq:NSE}) at $\Reynolds=400$ 
and $\Omega = 4\pi\!\times\!2\!\times\!16\pi$, originally identified in~\cite{Schneider2009} 
for $\Omega = 4\pi\!\times\!2\!\times\!8\pi$.
The solutions are localized in the spanwise $z$ direction and consist of 
two to three prominent pairs of alternating wavy roll-streak structures
embedded in a laminar background flow. 
Figure~\ref{fig:Re400fields}(a,b) is a traveling-wave solution $\butw$ of
(\ref{eq:NSE}) satisfying $[u,v,w](x,y,z,t)=[u,v,w](x-c_x t,y,z,0)$, where 
$c_x=0.028$ is the streamwise wavespeed. Figure~\ref{fig:Re400fields}(c,d) is a stationary, 
time-independent solution $\bueq$. The equilibrium $\bueq$ is symmetric under 
inversion $[u,v,w](x,y,z,t) = [-u,-v,-w](-x,-y,-z,t)$, and the traveling wave 
$\butw$ has a shift-reflect symmetry, $[u,v,w](x,y,z,t) = [u,v,-w](x+L_x/2,y,-z,t)$.
These symmetries ensure that neither $\bueq$ nor  $\butw$ drifts in the 
localization direction $z$.

\begin{figure}
\includegraphics[width=0.45\textwidth]{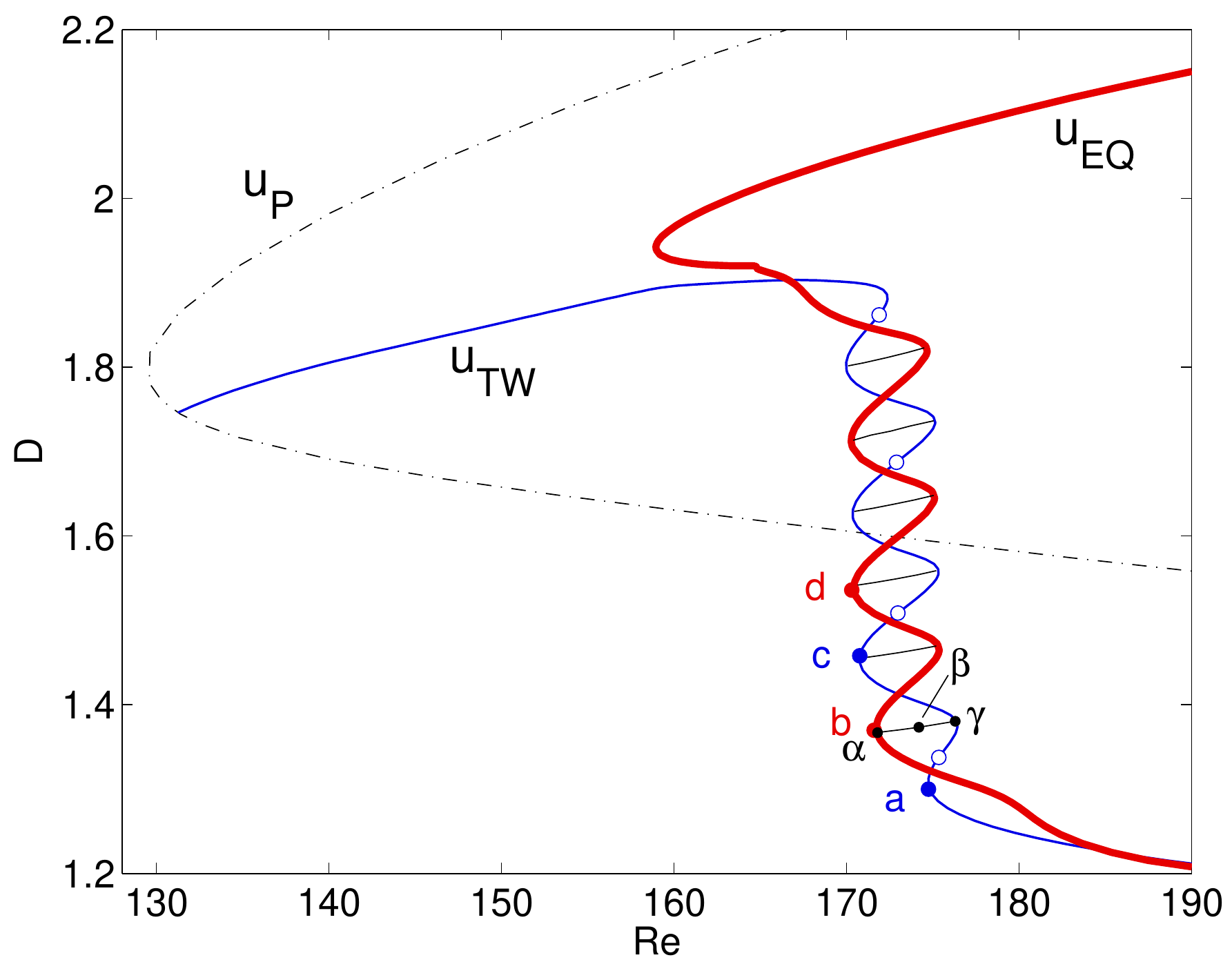}
\caption{\label{fig:snaking} (color online). Snaking of the
localized $\butw, \bueq$ solutions of plane Couette flow in $(\Reynolds,D)$ plane. 
The spatially periodic Nagata solution $\bu_{\text{P}}$ is shown as well; the $\butw$
solution connects with it near $(131, 1.75)$. Velocity fields of the 
localized solutions at the saddle-node bifurcations labeled \textit{a,b,c,d}
are shown in \reffig{fig:vfields}. The rung branches are shown with solid 
lines connecting the $\bueq$ and $\butw$ in the snaking region; velocity 
fields for the points marked $\alpha, \beta, \gamma$ are shown in \reffig{fig:rungs}.
Open dots on the $\butw$ traveling wave branch mark points at which the 
wavespeed passes through zero. 
} 
\end{figure}

\begin{figure*}
{\small (a)} \includegraphics[width=0.45\textwidth]{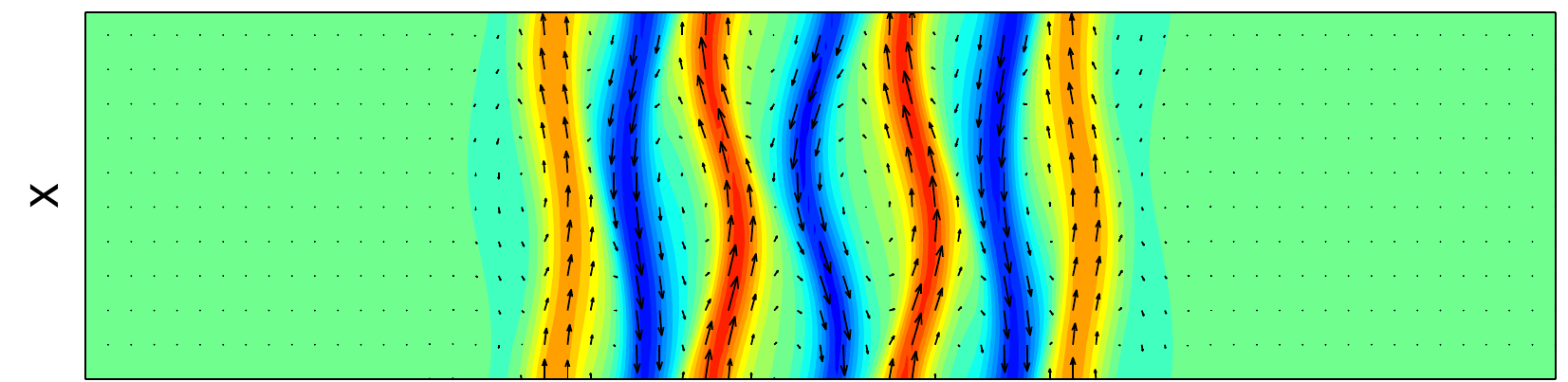} 
{\small (b)} \includegraphics[width=0.45\textwidth]{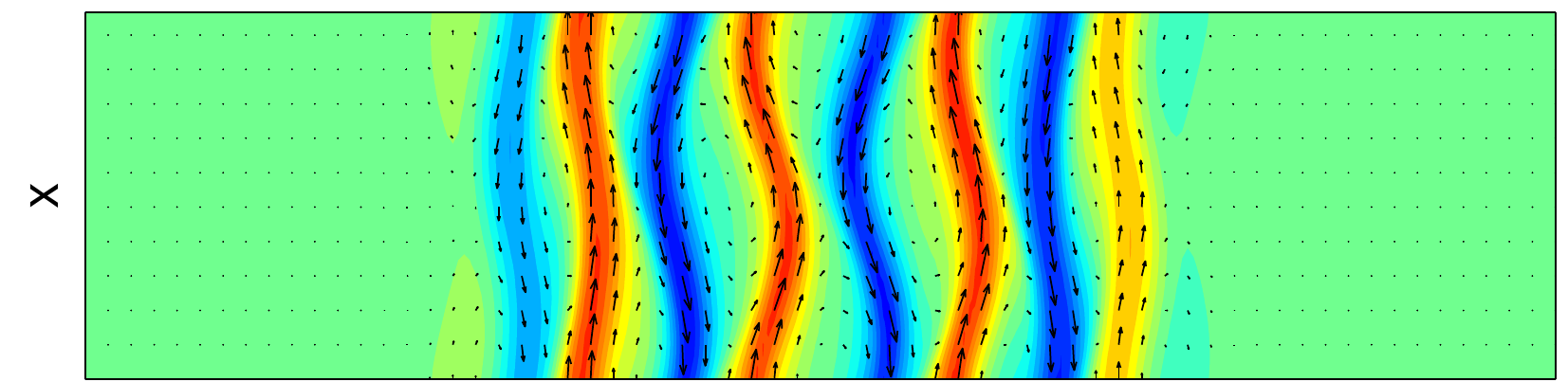} \\
{\small (c)} \includegraphics[width=0.45\textwidth]{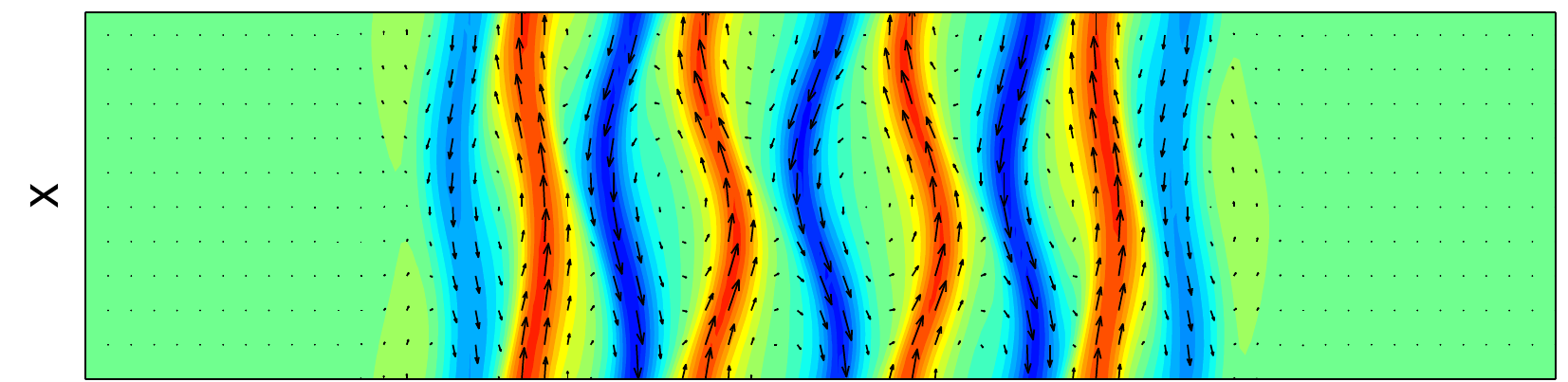} 
{\small (d)} \includegraphics[width=0.45\textwidth]{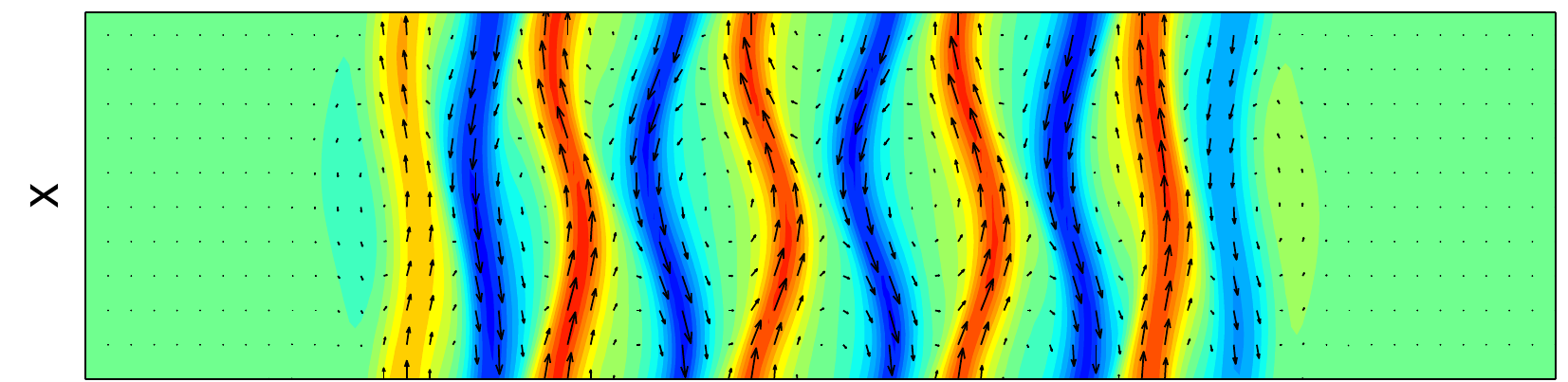} \\
{\small (e)} \includegraphics[width=0.45\textwidth]{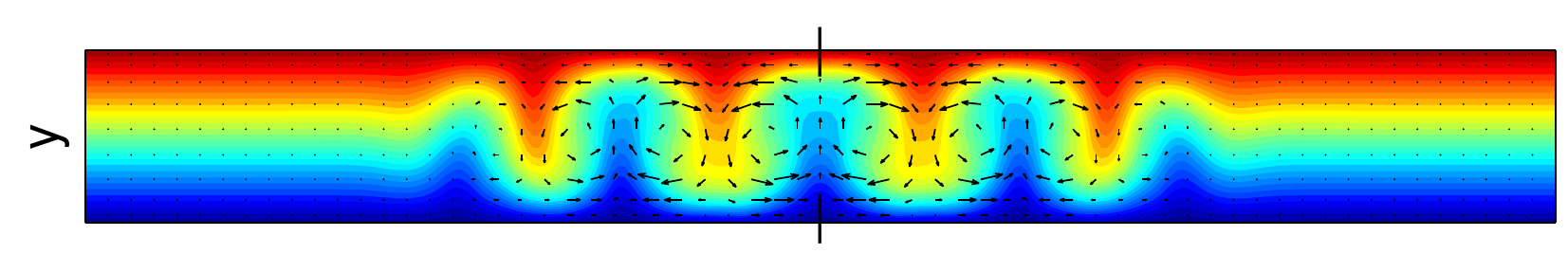} 
{\small (f)} \includegraphics[width=0.45\textwidth]{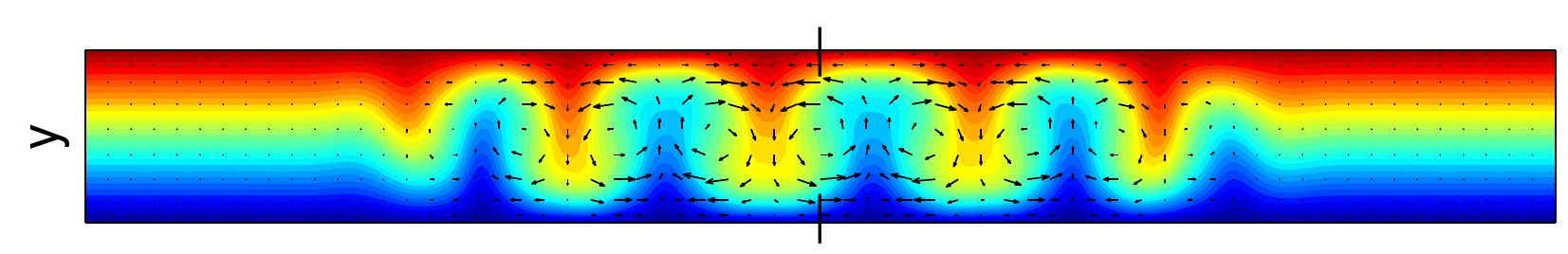} \\
{\small (g)} \includegraphics[width=0.45\textwidth]{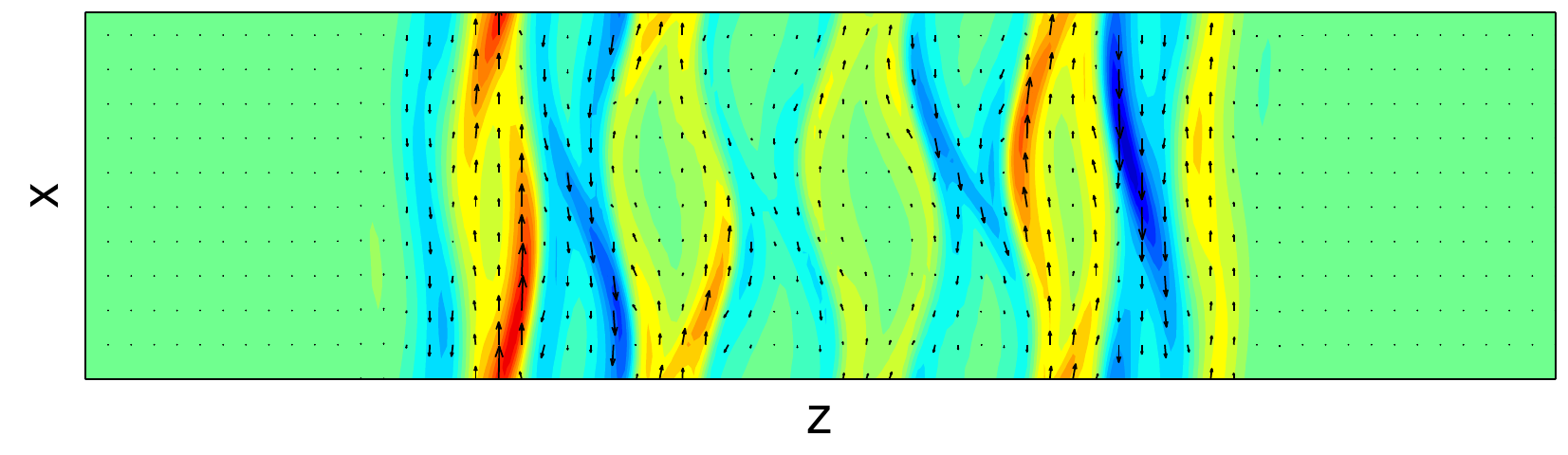} 
{\small (h)} \includegraphics[width=0.45\textwidth]{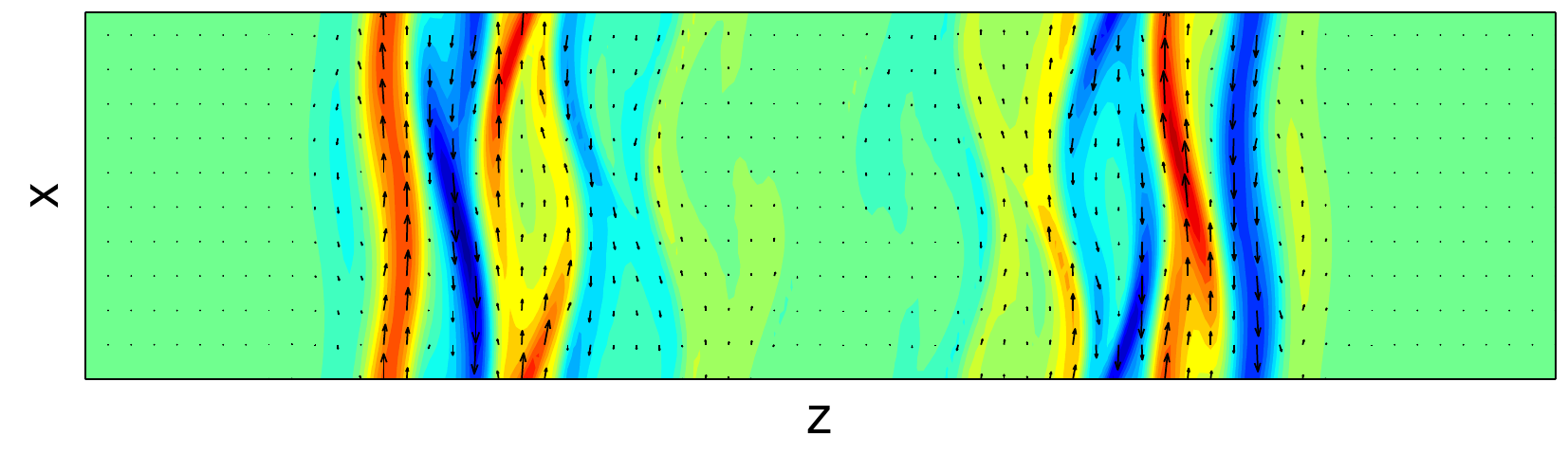}
\caption{\label{fig:vfields}
(color online). Localized traveling wave $\butw$ (left) and equilibrium $\bueq$ 
(right) solutions of plane Couette flow at points marked on the solution branches 
in \reffig{fig:snaking}. (a,c) show the velocity fields of $\butw$ at its first 
and second saddle-node  bifurcations, moving up each branch from lower to higher 
dissipation $D$; similarly (b,d) for $\bueq$. (e,f) show the $x$-averaged
velocity, with in-plane velocity indicated by arrows and streamwise velocity 
by the colormap. The marginal eigenfunctions at the saddle-node bifurcations 
(c,d) are shown in (g,h). 
}
\end{figure*}

\begin{figure}
{\small ($\alpha$)} \includegraphics[width=0.45\textwidth]{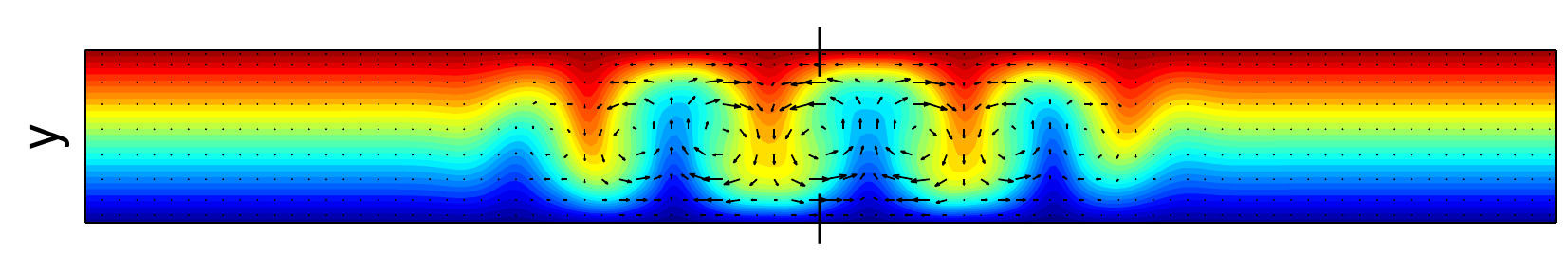} 
{\small ($\beta$)} \includegraphics[width=0.45\textwidth]{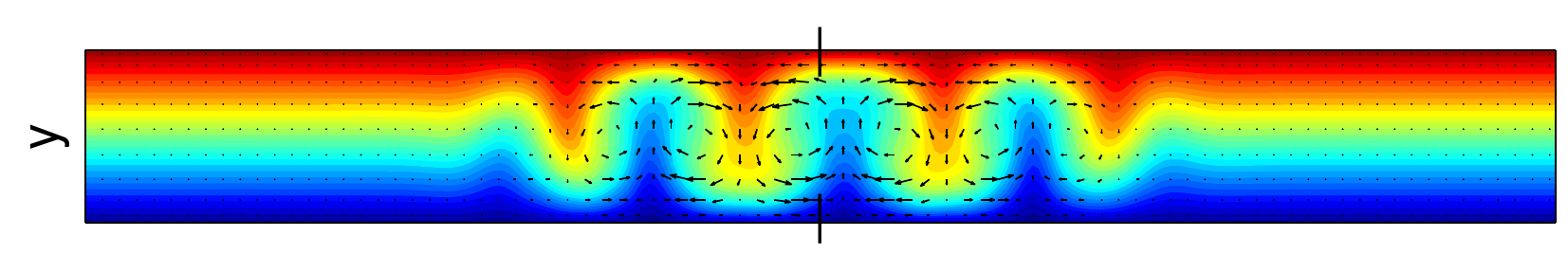} 
{\small ($\gamma$)} \includegraphics[width=0.45\textwidth]{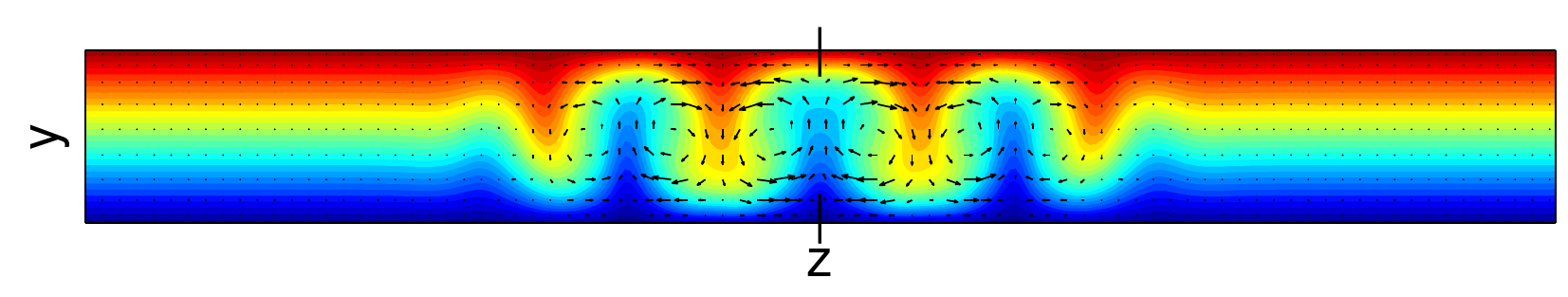} 
\caption{\label{fig:rungs}
(color online). Localized solutions of plane Couette flow along a rung branch,
for the points marked $\alpha, \beta, \gamma$ in \reffig{fig:snaking}, and 
plotted in terms of $x$-averaged streamwise velocity $\buxavg(y,z)$ as in 
\reffig{fig:Re400fields}(b,d). ($\alpha$) shows the beginning of the rung 
solution on the $\bueq$ branch with symmetry $\buxavg(y,z) = - \buxavg(-y,-z)$. 
Midway along the rung, ($\beta$) is non-symmetric. The rung terminates at
($\gamma$) on the $\butw$ branch with symmetry $\buxavg(y,z) = \buxavg(y,-z)$. 
}
\end{figure}

To continue these solutions in $\Reynolds$, we combine a Newton-Krylov 
hookstep algorithm~\cite{Viswanath2007} with quadratic extrapolation in 
pseudo-arclength along the solution branch. The Navier-Stokes equations are 
discretized with a Fourier-Chebyshev-tau scheme
in primitive variables and 3rd-order semi-implicit backwards differentiation 
time stepping. Bifurcations along the solution branches are characterized by 
linearized eigenvalues computed with Arnoldi iteration. The computations were 
performed with $32\!\times\!33\!\times\!256$ collocation points and $2/3$-style 
dealiasing and checked against computations with $(3/2)^3$ more gridpoints. 
The numerical software 
is available at~\url{www.channelflow.org}.


The bifurcation diagram in \reffig{fig:snaking} shows the $\butw$ and 
$\bueq$ solutions from Fig.~\ref{fig:Re400fields} under continuation in Reynolds 
number.
As $\Reynolds$ decreases below $180$, the solution branches
snake upwards in dissipation D; that is, they pass through a sequence of sub- 
and supercritical saddle-node bifurcations which nearly line up, creating 
a large multiplicity of localized solutions in $169 < \Reynolds < 177$. 
Each saddle-node bifurcation adds structure at the edges 
(fronts) of the localized solution while preserving its symmetry. This spatial 
growth is illustrated in \reffig{fig:vfields}, which shows the 
velocity fields at several points along the snaking branches. For example, 
\reffig{fig:vfields}(a) shows the $y=0$ midplane of $\butw$ at the saddle-node 
bifurcation marked \textit{a} in \reffig{fig:snaking}. Continuing up the 
solution branch from (a) to (c), the solution gains a pair of streaks at the fronts 
while the interior structure stays nearly constant. The marginal eigenfunction 
associated with the saddle-node bifurcation at (c) 
is shown in \refFig{fig:vfields}(g); it is weighted most heavily at the fronts of the 
localized solution and has the same symmetry, so that adding a small component of the 
eigenfunction strengthens and slightly widens the fronts, whereas subtraction weakens 
and shrinks them.

The spanwise wavelength of the interior structure of the localized 
solutions is approximately $\ell_z = 6.8 \approx 2\pi$. This value is 
selected by the fronts that connect the interior streaks to 
the laminar background, and it does not seem to vary across 
the snaking region or when compared between the two branches. 
The streamwise wavespeed $c_x$ of the traveling wave solution 
varies along the branch and in fact changes sign several times.
Points at which $c_x=0$ are marked in \reffig{fig:snaking} with open 
circles. The point marked \textit{a} has $c_x = 0.0062$. Rotation about 
the $z$ axis generates symmetric partners for both $\butw$ and $\bueq$;
for the former this results in a streamwise drift in the opposite direction. 
Thus the $\butw$ and $\bueq$ curves in~\reffig{fig:snaking} represent
four solution branches. The lower branches of both can be 
continued upwards in Reynolds number past $\Reynolds = 1000$.

Figure~\ref{fig:snaking} also shows six rungs of non-symmetric exact
localized solutions. These bifurcate from the snaking branches close
to the saddle nodes and are associated with marginal eigenfunctions
whose symmetry does not match the base state. Each rung connects to
both the $\bueq$ and the $\butw$ branch so solutions along the rungs 
smoothly interpolate between the two symmetry subspaces, as illustrated in 
\reffig{fig:rungs}. The rung solutions travel in $z$ as well as $x$, 
but with $z$ wavespeed three orders of magnitude smaller than $c_x$.

Due to the finite extent of the domain, the structures cannot grow
indefinitely, and the snaking behavior must terminate. As in other 
problems of this type, the details of this termination depend on a 
commensurability condition between the spanwise wavelength of the 
streaks within the localized solutions and the spanwise 
domain~\cite{Bergeon:2008p420}. 
At $L_z = 16 \pi$, $\butw$ connects at $(\Reynolds,D) = (131, 1.75)$ to the 
spatially periodic Nagata equilibrium with wavelength $\ell_z=2\pi$ .
The $\bueq$ branch does not appear to connect to any periodic solution. 
Instead, when this 
branch exits the snaking region its velocity field contains a localized 
defect that persists under continuation up to at least $\Reynolds=300$. 
At other values of $L_z$, both branches might either not connect to a 
periodic solution or connect to a different periodic solution.
For example, in Ref.~\cite{Schneider2009} it was shown that 
at $L_z = 8 \pi$ both the $\butw$ and $\bueq$ branches terminate on a 
branch of spatially periodic solutions, though that choice of $L_z$ was 
too narrow to allow the snaking structure to develop.


We have shown that homoclinic snaking in wide plane Couette channels
gives rise to a family of exact localized solutions with internal
structure similar to the periodic Nagata equilibrium. Thus, as
recently speculated~\cite{Knobloch:2008p529,Schneider2009}, the
localization mechanism studied in the SHE carries over to linearly 
stable shear flows, where it cannot be associated with an instability
of the uniform state~\cite{Romanov1973}. 
Physically, the localized states studied above consist of fronts
pinned to the periodic Nagata equilibrium. The periodic structure is
formed by pairs of counter-rotating, streamwise-oriented 
roll-streak structures, which also characterize other exact solutions 
linked to transitional turbulence in small domains. Therefore, localized
versions of the other known exact solutions should also exist and,
together with their heteroclinic connections, support localized
turbulence. In this sense the localized solutions studied here are a
first step towards generalizing the dynamical systems picture for
turbulence to extended flows.

Turbulent spots and stripes that are tilted against the flow
direction suggest the existence of exact solutions localized in both
spanwise and streamwise direction. Although a theory for localization
in two spatial dimensions is not yet available, numerical studies of
the SHE show that snaking does carry over to solutions localized in
two dimensions~\cite{LLoydSIADS08}. 
In PCF it is however not known if the same mechanism
also generates fully localized exact solutions, because a bifurcation
analysis would first require a fully localized solution to start the
continuation.
Such a solution is unfortunately not yet available. Nevertheless edge 
calculations both in pipe flow~\cite{Willis2009, Mellibovsky2009} and 
in extended plane Couette cells~\cite{Marinc2008,Schneider2009,Duguet2009} 
yield localized structures that show very mild dynamic fluctuations,
which suggests the existence of simple underlying fully localized
exact solutions.

\begin{acknowledgments}
We would like to thank Edgar Knobloch for helpful discussions and Predrag 
Cvitanovi\'c and Bruno Eckhardt for helpful comments on the manuscript.
T.M.S.\ was supported by German Research Foundation grant Schn 1167/1.
J.F.G.\ was supported by NSF grant DMS-0807574. J.B.\ was supported by 
NSF grant DMS-0602204. 
\end{acknowledgments}


\end{document}